\documentclass{amsart}

\newtheorem{theorem}{Theorem}[section]

\theoremstyle{definition}

\theoremstyle{remark}

\numberwithin{equation}{section}

\begin{document}



\title{A polynomial time algorithm for 3-SAT}

\author{Sergey Gubin }


\subjclass[2000]{Primary 68Q15, 68R10, 90C57}



\keywords{Computational complexity, Algorithms, Satisfiability, SAT, 3SAT}

\begin{abstract}
Article describes a class of efficient algorithms for 3SAT.
\end{abstract}

\maketitle

\section*{Introduction}
\label{s:Introduction}

This paper presents the compatibility matrix method for 3SAT. The compatibility matrix is an encoding of 3SAT instances using the contradictions between their clauses. This encoding is the first major result of this work.
\newline\indent
The compatibility matrix encoding allows parallel testing of all guesses. This paper describes several algorithms which perform the parallel testing of all guesses in polynomial time indeed. The algorithms transform the compatibility matrix into a similar encoding of the satisfying true-assignments. These efficient algorithms for 3SAT are the second major result of this work.
\newline\indent
Let us describe 3SAT as follows. Suppose, $f$ is a given conjunctive normal form (CNF) with clauses of length three or less. Let
\begin{equation}
\label{e:formula}
f = c_1 \wedge c_2 \wedge \ldots \wedge c_m
\end{equation}
In formula $f$, clauses $c_i$ are disjunctions of three or less non-complimentary literals (different Boolean variables or their negations):
\[
c_i \in \{\alpha \vee \beta \vee \gamma, ~ \alpha \vee \beta, ~ \alpha ~|~\alpha, \beta, \gamma \in \{b_j, \neg b_j ~|~  j = 1,2,\ldots, n \} \},
\]
- where $b_j$ are Boolean variables. Suppose, we have to decide whether there does exist such a true-assignment which would satisfy formula $f$:
\[
true  = f(b_1,b_2,\ldots,b_n)|_{b_j = \tau_j \in \{false, true \}}
\]
Such a decision problem is a 3SAT instance.
\newline\indent
Cook proved that 3SAT is a NP-complete problem \cite{cook0}. So, the whole history of the NP-completeness theory is the history of 3SAT. As a matter of fact, 3SAT was one of the first four NP-complete problems discovered \cite{cook0}. One might say that 3SAT is the most beautiful NP-complete problem discovered.

\section{Compatibility matrix}
\label{s:CompatibilityMatrix}

Let $T_i$ be the truth-table for clause $c_i$, $i = 1,2,\ldots,m$. Let's arbitrarily enumerate the true-assignments for the arguments of $c_i$ and let's write $T_i$ in the following form:
\[
\begin{array}{|c|c|c|}
\hline
T_i& \mbox{True-assignments} & c_i \\
\hline
1& \mbox{1-st true-assignment } & \mbox{Value of } c_i \mbox{ on the 1-st true-assignment}\\
2& \mbox{2-nd true-assignment } & \mbox{Value of } c_i \mbox{ on the 2-nd true-assignment}\\
\vdots & \vdots & \vdots \\
\hline
\end{array}
\]
\indent
Let $S_{i\mu}$ be $\mu$-th string from truth-table $T_i$. By definition, any two strings $S_{i\mu}$ and $S_{j\nu}$ are compatible if the following two conditions are satisfied:
\begin{description}
\item[Condition 1]
The $\mu$-th value of $c_i$ is $true$ and the $\nu$-th value of $c_j$ is $true$;
\item[Condition 2]
The $\mu$-th true-assignment in table $T_i$ and the $\nu$-th true- assignment in table $T_j$ do not contradict each other. In other words, if clauses $c_i$ and $c_j$ share a variable, then this variable has to have the same true- assignment in both strings $S_{i\mu}$ and $S_{j\nu}$.
\end{description}
When at least one of these conditions is not satisfied, then the strings are incompatible. Let us emphasize, $i$ and $j$ may be equal in this definition, i.e. strings $S_{i\mu}$ and $S_{j\nu}$ may be from the same truth-table.
\newline\indent
Let's build a compatibility box for each clause couple $(c_i,c_j)$, $i,j = 1,2,\ldots,m$. Let $k_i$ and $k_j$ be the lengths of clauses $c_i$ and $c_j$, appropriately. Then, the compatibility box for clauses $c_i$ and $c_j$ is a Boolean matrix\footnote{Boolean matrix is a matrix whose elements are $true$ or $false$.} $C_{ij} = (x_{\mu\nu})_{2^{k_i} \times 2^{k_j}}$ with the following elements:
\[
x_{\mu\nu} = \left \{ \begin{array}{cl}
true, & \mbox{Strings} ~ S_{i\mu} ~ \mbox{and} ~ S_{j\nu} ~ \mbox{are compatible} \\
false, & \mbox{Otherwise} \\
\end{array} \right.
\]
Having all $m^2$ compatibility boxes built, we can aggregate them in a box matrix $C$:
\[
C = (C_{ij})_{m \times m}
\]
- where the size is given in ``boxes''. Matrix $C$ is a Boolean box matrix. For 3SAT, size of $C$ is
\[
(\sum_{i = 1}^m 2^{k_i}) \times (\sum_{i = 1}^m 2^{k_i}) \leq (8m) \times (8m) = O(m \times m),
\]
- where $k_i$ is length of clause $c_i$. We call matrix $C$ a compatibility matrix for formula \ref{e:formula}.
\newline\indent
Let us notice that the compatibility matrix is a symmetric matrix:
\begin{equation}
\label{e:symmetry}
C = C^T, ~ C_{ij} = C_{ji}^T.
\end{equation}
Also, let us notice that, due to Condition 2 for the string compatibility, diagonal boxes $C_{ii}$ are diagonal matrices, i.e. all non-diagonal elements in these boxes are $false$:
\begin{equation}
\label{e:diagonal}
C_{ii} = \left ( \begin{array}{ccc}
t_1 & false & \ldots  \\
false & t_2 & \ddots \\
\vdots & \ddots & \ddots  \\
\end{array} \right ), ~ t_\mu \in \{true, false\}
\end{equation}
Because clauses in formula \ref{e:formula} are disjunctions, there is the only one value $false$ among values $t_\mu$ in matrix \ref{e:diagonal}.
\newline\indent
As usual, we can use a $(0,1)$-version of the compatibility matrix. In the version of matrix $C$, values $true$ are replaced with $1$, and values $false$ are replaced with $0$.

\section{Solution grid and depletion}
\label{s:SolutionGrid}

Traditionally, guessing in 3SAT is the checking whether a true-assignment satisfies formula \ref{e:formula}. But, any true-assignment for the variables in formula \ref{e:formula} contains/consists of the true-assignments for clauses $c_i$. So, any true-assignment produces in the compatibility matrix a grid of elements, one element per compatibility box:
\[
\gamma = \{x_{\mu\nu ij}~|~\mu = \mu(i), ~ \nu = \nu(j),~i,j=1,2,\ldots,m \},
\]
- where $x_{\mu\nu ij}$ is the $(\mu,\nu)$-element of compatibility box $C_{ij}$, and $\gamma$ is a grid of such elements in compatibility matrix $C$, one element per compatibility box.
\newline\indent
Based on the grid presentation, we can state that any true-assignment satisfying formula \ref{e:formula} is presented in compatibility matrix $C$ with a grid of elements, one element per compatibility box, which all are $true$. And visa versa, if there is such a grid of elements in the compatibility matrix, one element per compatibility box, whose all elements are $true$, then formula \ref{e:formula} is satisfiable. We call such a grid which consists of all $true$-elements\footnote{$True$-element is an element which is equal $true$, and $false$-element is an element which is equal $false$.}, one element per compatibility box, a solution grid.
\newline\indent
Based on property \ref{e:diagonal} of the compatibility matrix, we could even remove the ``one-element-per-box'' requirement from the solution grid definition.
\begin{theorem}
\label{t:grid}
Formula \ref{e:formula} is satisfiable iff compatibility matrix for that formula contains a solution grid.
\end{theorem}
Solution grid is a pattern of satisfiability in the compatibility matrix encoding of formula \ref{e:formula}. And theorem \ref{t:grid} reduces the guessing to the searching the matrix for the pattern. In the search, any heuristic is welcome.
\newline\indent
Our method is a detection of those elements of the compatibility matrix which do not belong to any solution grid and the assigning to them value $false$ or $0$ depending on the presentation of the compatibility matrix. We call the replacement of a $true$-element with value $false$ a depletion of the compatibility matrix.
\begin{theorem}
\label{t:pattern}
Suppose, there is a filter which finds in the compatibility matrix for formula \ref{e:formula} all such ``$true$''-elements which do not belong to any solution grid. Suppose, all these elements were replaced with value ``$false$''. Then, formula \ref{e:formula} is satisfiable iff there is not any compatibility box filled with ``$false$'' entirely.
\end{theorem}
In fact, for unsatisfiable formula \ref{e:formula}, the matrix emerging from the depletion described in theorem \ref{t:pattern} will be filled with $false$ entirely.
\newline\indent
A compatibility box entirely filled with $false$ we call a pattern of unsatisfiability. One might say that our method is a search of the compatibility matrix for the pattern. For the method, it does not matter what particular filter is used for the depletion. For the filter, the only objective is to preserve at least one solution grid when the grids exist. Then, no information will be lost in the sense of computational complexity.

\section{Basic algorithm}
\label{s:basic}

The following algorithm is a depletion filter. We call it a basic algorithm:
\begin{description}
\item[Init]
Build the compatibility matrix for formula \ref{e:formula}.
\newline
Let $s = 1$. Let $C_s$ be the compatibility matrix. Let $C_{sij}$ be the compatibility boxes, where $i,j = 1,2,\ldots,m$.
\item[Depletion]
Let $s=s+1$. Calculate compatibility matrix $C_{s}$ as follows\footnote{Here, product of two Boolean matrices $A=(a_{\mu\nu})$ and $B=(b_{\mu\nu})$ of the appropriate sizes (the number of columns in $A$ has to be equal to the number of rows in $B$) is the following Boolean matrix:
\[
AB = (\bigvee_k a_{ik}\wedge b_{kj}).
\]
Conjunction of Boolean matrices of the same size is the matrix of conjunctions of the appropriate elements of the matrices.}:
\begin{equation}
\label{e:depletion}
C_{sij} = \bigwedge_{k=1}^m C_{s-1,ik}C_{s-1,kj}, 
\end{equation}
- where $C_{sij}$ is the $(i,j)$-th compatibility box of $C_s$, $i,j = 1,2,\ldots,m$. 
\newline\indent
Let $x_{s\mu\nu ij}$ be the $(\mu\nu)$-th element of compatibility box $C_{sij}$. Then, due to formula \ref{e:depletion}, the element is 
\begin{equation}
\label{e:multi}
x_{s\mu\nu ij} = \bigwedge_{k=1}^m (\bigvee_{1 \leq \alpha \leq 2^3} x_{s-1,\mu\alpha ik} \wedge x_{s-1,\alpha\nu kj}).
\end{equation}
Due to property \ref{e:symmetry} of the compatibility matrix, the last expression can be rewritten in any of the following ways:
\begin{equation}
\label{e:element}
x_{s\mu\nu ij} = \bigwedge_k (\bigvee_\alpha x_{s-1,\alpha\mu ki} \wedge x_{s-1,\alpha\nu kj}) = \bigwedge_k (\bigvee_\alpha x_{s-1,\mu\alpha ik} \wedge x_{s-1,\nu\alpha jk}).
\end{equation}
\item[Iterations]
If $C_{s} \neq C_{s-1}$, then go to the previous step. Otherwise, continue.
\item[Decision]
If $C_{s}$ is entirely filled with $false$, then formula \ref{e:formula} is unsatisfiable - decision ``NO''. Otherwise, formula \ref{e:formula} is satisfiable - decision ``YES''.
\end{description}
Obviously, when during an iteration the pattern of unsatisfiability arises (value $false$ fills a compatibility box entirely), then the algorithm can be stopped with decision ``NO''. Because, formula \ref{e:depletion} will propagate this value $false$ all over the compatibility matrix during the next two iterations, at most.
\newline\indent
For 3SAT, the computational complexity of the algorithm can be estimated as $O(m^5)$: there is $O(m^2)$ iterations (each iteration depletes at least one element from the compatibility matrix); there is $m^2$ depletions on each iteration; and each depletion takes time $O(m)$. 
\begin{theorem}
\label{t:basic}
Any ``$true$''-element in the matrix emerging after running of the basic algorithm belongs to a solution grid.
Formula \ref{e:formula} is unsatisfiable iff the basic algorithm produces the pattern of unsatisfiability for this formula.
\end{theorem}
\begin{proof}
Clauses in formula \ref{e:formula} can be expressed in their disjunctive forms (DF). Depending on their length, the clauses are
\begin{equation}
\label{e:triads}
\begin{array}{rl}
\alpha = & false ~\vee ~\alpha \\
\alpha \vee \beta = & false ~\vee ~\alpha \wedge \beta ~\vee ~\alpha \wedge \bar{\beta} ~\vee ~ \bar{\alpha} \wedge \beta \\
\alpha \vee \beta \vee \gamma = & false ~ \vee ~ \alpha \wedge \beta \wedge \gamma ~ \vee ~ \alpha \wedge \beta \wedge \bar{\gamma} ~ \vee ~ \alpha \wedge \bar{\beta}\wedge \gamma ~ \vee \\
\vee & \alpha \wedge \bar{\beta}\wedge \bar{\gamma} ~ \vee ~ \bar{\alpha} \wedge \beta \wedge \gamma ~ \vee ~ \bar{\alpha} \wedge \beta \wedge \bar{\gamma} ~ \vee ~ \bar{\alpha} \wedge \bar{\beta} \wedge \gamma \\
\end{array}
\end{equation}
- where $\alpha$, $\beta$, and $\gamma$ are literals. 
\newline\indent
Let's call all conjunctions in DF \ref{e:triads} the triads (value $false$ in the formulas is a triad, too). For each clause, there is a natural one-to-one relation between the clause's triads and true-assignments: that true-assignment which makes a clause equal $false$ relates to triad $false$; the rest of the true-assignments are in the relation with those triads to which they deliver value $true$. Let's enumerate the triads in DF \ref{e:triads} using that one-to-one relations: triad $\tau_{\mu i}$ is that triad which relates to $\mu$-th true-assignment for clause $c_i$. 
\newline\indent
Let's replace clauses in formula \ref{e:formula} by their DF \ref{e:triads}, and let's open all parentheses in the formula. The result of that is the following DF:
\[
f = \bigvee_{\mu_1, \mu_2, \ldots, \mu_m} ~ \bigwedge_{i=1}^m \tau_{\mu_i i} 
\]
- where the disjunction is taken over all appropriate combinations of indexes $\mu_i$. Then, formula \ref{e:formula} is satisfiable iff there is at least one satisfiable conjunction 
\begin{equation}
\label{e:gamma}
\Gamma(\mu_1, \mu_2, \ldots, \mu_m) = \bigwedge_{i=1}^m \tau_{\mu_i i}
\end{equation}
In other words, formula \ref{e:formula} is satisfiable iff there is at least one conjunction \ref{e:gamma} without complimentary literals.
\newline\indent
Now, let's write the following symbolic box matrix $Y$: 
\[
Y = (Y_{ij})_{m\times m},
\]
- where $Y_{ij}$ are the boxes. The symbolic boxes in matrix $Y$ are as follows:
\[
Y_{ij} = (\tau_{\mu i} \wedge \tau_{\nu j})_{2^{k_i} \times 2^{k_j}},
\]
- where $k_i$ and $k_j$ are lengths of clauses $c_i$ and $c_j$ appropriately.
\newline\indent
Let's calculate powers of matrix $Y$ using formulas \ref{e:depletion} and \ref{e:multi}:
\[
Y^1 = Y, ~ Y^{s+1} = (\bigwedge_{k=1}^m Y_{s,ik}Y_{s,kj})_{m\times m}, ~s \geq 1
\]
- where boxes $Y_{s,*,*}$ are the appropriate boxes in box matrix $Y^{s}$. Due to formula \ref{e:multi}, elements of matrix $Y^s = (y_{\alpha\beta})_{m\times m}$ are some DF of the following kind:
\begin{equation}
\label{e:y}
y_{\alpha\beta} = \bigvee_{\sigma_1,\sigma_2,\ldots} \tau_{\mu_{\sigma_1},\sigma_1} \wedge \tau_{\mu_{\sigma_2},\sigma_2} \wedge \ldots
\end{equation}
- where $(\sigma_1,\sigma_2,\ldots)$ are all appropriate combinations of the triads' indexes. 
\newline\indent
The most important for us is to notice that, starting from some power $s = O(m^2)$ of matrix $Y$, each conjunction in DF \ref{e:y} contains at least $m$ different triads. But, any conjunction of more than $m$ different triads contains at least two different triads from the same clause. But, any such conjunction equals $false$ because there are complimentary literals in those different triads from the same clause. Thus, starting from that power, all DF \ref{e:y} get finalized as follows:
\begin{equation}
\label{e:final}
y_{\alpha\beta} = \bigvee_{\mu_1, \mu_2, \ldots, \mu_m} \Gamma(\mu_1, \mu_2, \ldots, \mu_m)
\end{equation}
- where the disjunction is taken over all appropriate combinations of indexes $\mu_i$. 
\newline\indent
Let's rewrite $\Gamma(\mu_1, \mu_2, \ldots, \mu_m)$ as follows:
\[
\Gamma(\mu_1, \mu_2, \ldots, \mu_m) = \bigwedge_{i=1}^m \tau_{\mu_i i} = 
\]
\[
\begin{array}{ccc}
= & (\tau_{\mu_1,1} \wedge \tau_{\mu_1,1}) ~ \wedge ~ (\tau_{\mu_1,1} \wedge \tau_{\mu_2,2}) ~ \wedge ~ \ldots ~\wedge ~ (\tau_{\mu_1,1} \wedge \tau_{\mu_m,m}) & \wedge \\
\wedge & (\tau_{\mu_2,2} \wedge \tau_{\mu_1,1}) ~ \wedge ~ (\tau_{\mu_2,2} \wedge \tau_{\mu_2,2}) ~ \wedge ~ \ldots ~ \wedge ~ (\tau_{\mu_2,2} \wedge \tau_{\mu_m,m}) & \wedge \\
\wedge & \ldots & \wedge \\
\wedge & (\tau_{\mu_m,m} \wedge \tau_{\mu_1,1}) \wedge (\tau_{\mu_m,m} \wedge \tau_{\mu_2,2}) \wedge ~ \ldots ~ \wedge (\tau_{\mu_m,m} \wedge \tau_{\mu_m,m}) & \\
\end{array}
\]
Each parenthesis in the last presentation is an element of the compatibility matrix for formula \ref{e:formula}:
\[
\tau_{\mu i} \wedge \tau_{\nu j} = x_{\mu\nu ij}
\]
- where $x_{\mu\nu ij}$ is the $(\mu,\nu)$-the element in compatibility box $C_{ij}$. Element $x_{\mu\nu ij}$ is $true$ iff triads $\tau_{\mu i}$ and $\tau_{\nu j}$ contain no complimentary literals. Thus, conjunctions \ref{e:gamma} are associated with the grids of the compatibility matrix's elements, one element per compatibility box:
\begin{equation}
\label{e:express}
\begin{array}{rcc}
\Gamma(\mu_1, \mu_2, \ldots, \mu_m) = & x_{\mu_1,\mu_2, 1,1} ~ \wedge ~ x_{\mu_1,\mu_2,1,2} ~ \wedge ~ \ldots ~\wedge ~ x_{\mu_1,\mu_m,1,m} & \wedge \\
\wedge & x_{\mu_2,\mu_1, 2,1} ~ \wedge ~ x_{\mu_2,\mu_2,2,2} ~ \wedge ~ \ldots ~\wedge ~ x_{\mu_2,\mu_m,2,m} & \wedge \\
\wedge & \ldots & \wedge \\
\wedge & x_{\mu_m,\mu_1, m,1}  \wedge  x_{\mu_m,\mu_2,m,2}  \wedge ~ \ldots ~\wedge  x_{\mu_m,\mu_m,m,m} &  \\
\end{array}
\end{equation}
Among all conjunctions \ref{e:gamma}, only those are equal $true$ whose elements constitute a solution grid in the compatibility matrix:
\[
\begin{array}{rcl}
\Gamma(\mu_1, \mu_2, \ldots, \mu_m) = & x_{\mu_1,\mu_2, 1,1} ~ = ~ x_{\mu_1,\mu_2,1,2} ~ = ~ \ldots ~ = ~ x_{\mu_1,\mu_m,1,m} & = \\
= & x_{\mu_2,\mu_1, 2,1} ~ = ~ x_{\mu_2,\mu_2,2,2} ~ = ~ \ldots ~ = ~ x_{\mu_2,\mu_m,2,m} & = \\
= & \ldots & = \\
= & x_{\mu_m,\mu_1, m,1}  =  x_{\mu_m,\mu_2,m,2}  = ~ \ldots ~ =  x_{\mu_m,\mu_m,m,m} & = ~ true \\
\end{array}
\]
\end{proof}
Compatibility matrix is an encoding of formula \ref{e:formula}. Theorem \ref{t:basic} shows that the basic algorithm transforms this encoding into an encoding of those true-assignments which satisfy the formula. Each $true$-element in the resulting matrix belongs to a solution grid. The true-assignments appropriate to any resulting $true$-element are a part of a satisfying true-assignment. These parts can be used to compute the whole satisfying true-assignments by the method of self-reducibility.
\newline\indent
Let us notice that, due to formula \ref{e:multi}, the length of conjunctions in DF \ref{e:y} grows exponentially over power of matrix $Y$. So, the expected number of iterations in the basic algorithm can be estimated as $O(\log m)$. Thus, the expected running time of the algorithm can be estimated as $O(m^3\log m)$.
\newline\indent
Also, in the case of unsatisfiable formula \ref{e:formula}, number $m$ in all estimations of time can be replaced  by the number of clauses in the minimal unsatisfiable sub-formula of formula \ref{e:formula}. That is due to the conjunctions in the formula \ref{e:depletion}: the basic algorithm restricted to those compatibility boxes which are built for a minimal unsatisfiable sub-formula will deplete less $true$-elements but it will produce the pattern of unsatisfiability.
\newline\indent
Before going further, let us notice that, due to Condition 1 of the strings compatibility, the true-assignments, which deliver to their clauses value $false$, can be missed when building the compatibility matrix, because they will produce in their compatibility boxes the rows and columns entirely filled with $false$. So, the compatibility matrix can be built solely on the contradictions between true-assignments satisfying different clauses separately.

\section{Asynchronous depletion}
\label{s:asynch}

Obviously, depletions \ref{e:depletion} could use the current values of the compatibility boxes which were computed during the current iteration. Such an asynchronous depletion can be expressed with the following back-feeds:
\begin{equation}
\label{e:asynch}
C_{ij} = C_{ij} \wedge C_{ik} C_{kj}, ~ 1 \leq i, j, k \leq m
\end{equation}
\indent
Really, diagonal boxes $C_{ii}$ in the compatibility matrix are the diagonal Boolean matrices, and\footnote{$false = false, ~ false < true, ~ true = true$.}
\[
C_{ij} = C_{ii}C_{ij} = C_{ij}C_{jj}, ~ C_{ii} \leq C_{ij}C{ij}^T, ~ C_{jj} \leq C_{ij}^T C_{ij}
\]
Thus, depletions \ref{e:depletion} are conjunctions of depletions \ref{e:asynch}. 
\newline\indent
Then, the basic algorithm can be reformulated as follows:
\begin{description}
\item[Init]
Compose the compatibility boxes for formula \ref{e:formula} and select a schema to iterate index-triplets 
\begin{equation}
\label{e:triplets}
(i,k,j): ~ 1 \leq i, j, k \leq m
\end{equation} 
The schemes can be context-sensitive and have other specifics. For example, a schema could ``visit'' the triplets with different frequencies or even miss some of them. The simplest schemes just cycle trough all index-triplets;
\item[Depletions]
Follow the selected iteration schema and deplete the appropriate compatibility boxes using transformations \ref{e:asynch} until each of these boxes became an invariant under the depletions.
\item[Decision]
If the pattern of unsatisfiability emerges during the depletions, then formula \ref{e:formula} is unsatisfiable. Otherwise, the formula is satisfiable.
\end{description} 
The freedom of selection the iteration schema creates a family of depletion filters: the asynchronous algorithms differ one from another by their iteration schemes. Theorem \ref{t:basic} implies 
\begin{theorem}
\label{t:asynch}
Let an iteration schema traverse all index-triplets \ref{e:triplets}. Then, formula \ref{e:formula} is unsatisfiable iff depletions \ref{e:asynch} produce the pattern of unsatisfiability for this formula.
\end{theorem}
For 3SAT, computational complexity of the asynchronous algorithms can be estimated as $O(m^3)$: there are $m^3$ index-triplets \ref{e:triplets}; during iterations \ref{e:asynch}, each of these triplets involves $O(1)$ Boolean operations to deplete the appropriate compatibility boxes of size $2^3 \times 2^3$ or less.
\newline\indent
Obviously, depletions \ref{e:asynch} will detect the pattern of unsatisfiability (if there is one) in time $O(s^3)$, where $s$ is the number of clauses in the shortest unsatisfiable sub-formulas of formula \ref{e:formula}.

\section{Triangular algorithms}
\label{s:triangular}

Due to the symmetry of the compatibility matrix, any asynchronous depletion can be restricted to those compatibility boxes which are located over/under the major diagonal of the matrix. For example, the compatibility boxes allocated over the major diagonal of the compatibility matrix are described with the following index-triplets: 
\begin{equation}
\label{e:indexes}
1 \leq i < k < j \leq m
\end{equation}
For these index-triplets, depletions \ref{e:asynch} have to be rewritten as follows:
\begin{equation}
\label{e:triangular}
\left \{ \begin{array}{r}
C_{ij} = C_{ij} \wedge C_{ik} C_{kj} \\
C_{ik} = C_{ik} \wedge C_{ij} C_{kj}^T \\
C_{kj} = C_{kj} \wedge C_{ik}^T C_{ij} \\
\end{array} \right.
\end{equation}
Theorem \ref{t:asynch} implies the following.
\begin{theorem}
\label{t:triangular}
Let an iteration schema traverse all index-triplets \ref{e:indexes}. Then, formula \ref{e:formula} is unsatisfiable iff depletions \ref{e:triangular} produce the pattern of unsatisfiability for this formula.
\end{theorem}
Any such algorithm we call a triangular algorithm.

\section{Square algorithms}
\label{s:square}

For the $(0,1)$-version of the compatibility matrix, depletions \ref{e:element} can be rewritten ``sub-tropically'':
\begin{equation}
\label{e:square}
x_{\mu\nu ij} = \min \{x_{\mu\nu ij},  \min \{ 1,\left\lfloor  \frac{1}{m}\sum_{\beta=1}^m \max \{ x_{\mu\alpha i\beta} x_{\alpha\mu \beta j}~|~1 \leq \alpha \leq 2^{k_\beta} \}\right\rfloor \} \},
\end{equation}
- where $x_{\mu\nu ij}$ is the $(\mu,\nu)$-th element of compatibility box $C_{ij}$, $k_\beta$ is the number of literals in clause $c_\beta$, and $\left\lfloor * \right\rfloor$ is the whole part of a number. 
\newline\indent
Obviously, all the above algorithms do work with depletions \ref{e:square} and have the same computational complexity. The above theorems imply
\begin{theorem}
\label{t:square}
Formula \ref{e:formula} is unsatisfiable iff a square algorithm produces the pattern of unsatisfiability for this formula.
\end{theorem}
We call such algorithms the square algorithms because of the following reason.
\newline\indent
Obviously, the compatibility matrix for formula \ref{e:formula} contains a solution grid iff there are $m$ such permutation matrices $X_1$, $X_2$, $\ldots$, and $X_m$ which would satisfy the following system of inequalities\footnote{For matrices $A=(a_{\mu\nu})$ and $B=(b_{\mu\nu})$ of the same size, relation $A \geq B$ means that $\forall \mu,\nu~(a_{\mu\nu} \geq b_{\mu\nu})$.}:
\begin{equation}
\label{e:permut}
X_i C_{ij} X_j^T \geq \left ( \begin{array}{ccc}
1&0&\ldots \\
0&0&\ldots \\
\vdots&\vdots&\ddots \\
\end{array} \right )_{2^{k_i} \times 2^{k_j}}, ~ i,j = 1,2,\ldots,m
\end{equation}
- where $k_i$ and $k_j$ are the numbers of literals in clauses $c_i$ and $c_j$ appropriately. This system just reflect that fact that, for a satisfiable formula \ref{e:formula}, there is such a re-enumeration of true-assignments in truth-tables $T_i$ which would move a solution grid's elements in the upper left corners of their compatibility boxes.
\newline\indent
System \ref{e:permut} is a system of integral bilinear inequalities. Iterations \ref{e:square} just solve this system. Formula \ref{e:square} iteratively compares the squares of the compatibility matrix with the squares of that matrix on the right side of system \ref{e:permut} and depletes the compatibility matrix appropriately.
\newline\indent
The result of such depletions is a general solution of system \ref{e:permut} in the form of  the compatibility matrix encoding of all true-assignments satisfying formula \ref{e:formula}.

\section*{Conclusion}

This paper described on high-level several polynomial time algorithms for 3SAT. The algorithms are based on a original encoding of 3SAT instances - a compatibility matrix. The encoding reflects the contradictions between true-assignments satisfying different clauses separately. Such a ``hologram'' of 3SAT instances allows parallel testing of all guesses. The algorithms transform the compatibility matrix into a matrix encoding of the satisfying true-assignments.
\newline\indent
Basically, the compatibility matrix presents 3SAT instances in the form a jigsaw puzzle, and the algorithms simulate the human activities during the solving of the puzzle. 
\newline\indent
For some of the algorithms, there are demos at \cite{demos}. Readers can use these demos to probe the algorithms with their own 3SAT instances.

\end{document}